\documentclass[sigconf, screen, nonacm]{acmart}

\usepackage{algorithmic}
\usepackage{textcomp}
\usepackage{comment}
\usepackage{booktabs}
\usepackage{array}
\usepackage{tabularx}
\usepackage{enumitem}
\usepackage{xurl}

\newcommand{\system}{Lodestone}
\makeatletter
\AtBeginDocument{%
  \fancypagestyle{standardpagestyle}{%
    \fancyhf{}%
    \renewcommand{\headrulewidth}{\z@}%
    \renewcommand{\footrulewidth}{\z@}%
    \fancyhead[L]{\@headfootfont\shorttitle}%
    \fancyhead[R]{\@headfootfont\@shortauthors}%
    \fancyfoot[C]{\if@ACM@printfolios\footnotesize\thepage\fi}%
  }%
  \pagestyle{standardpagestyle}%
}
\makeatother

\begin{document}

%%%%%%%%%%%---SETME-----%%%%%%%%%%%%%
\title{\system{}: Automated Semantic Discovery of Datacenter Hardware Specialization}

\makeatletter
\newcommand{\paperNames}{}
\newcommand{\paperAuthor}[3][]{%
  \author{#2}%
  \affiliation{\institution{#3}}%
  \g@addto@macro\paperNames{#2#1\enspace}%
}
\newcommand{\utmark}{\textsuperscript{1}}
\newcommand{\googlemark}{\textsuperscript{2}}

\paperAuthor[\utmark]{Peizhe Liu}{The University of Texas at Austin}
\paperAuthor[\googlemark]{Akanksha Jain}{Google}
\paperAuthor[\googlemark]{Gerasimos Gerogiannis}{Google}
\paperAuthor[\googlemark]{Vidushi Dadu}{Google}
\paperAuthor[\googlemark]{Parthasarathy Ranganathan}{Google}
\paperAuthor[\utmark]{Jovan Stojkovic}{The University of Texas at Austin}

\if@ACM@anonymous\else
  \renewcommand{\@mkauthors@iii}{%
    \global\setbox\mktitle@bx=\vbox{%
      \unvbox\mktitle@bx
      \vskip 13.5pt
      \hbox to \textwidth{\hfil
        {\normalfont\fontsize{10}{12}\selectfont\paperNames\unskip}%
        \hfil}%
      \vskip 3pt
      \hbox to \textwidth{\hfil
        {\@affiliationfont The University of Texas at Austin\utmark, Google\googlemark}%
        \hfil}%
      \bigskip}}
\fi
\makeatother

\renewcommand{\shortauthors}{P. Liu, A. Jain, G. Gerogiannis, V.  Dadu, P. Ranganathan, J. Stojkovic}

%%%%%%%%%%%%%%%%%%%%%%%%%%%%%%%%%%%%

%%%%%% -- PAPER CONTENT STARTS-- %%%%%%%%

\begin{abstract}
Hardware specialization 
via instruction set extensions or dedicated accelerators 
can improve datacenter efficiency. 
Unfortunately, identifying computations that justify dedicated hardware is difficult. 
Existing approaches analyze 
hot functions, 
instruction patterns, or known overheads such as datacenter tax.  
These approaches can miss fine-grained computations whose aggregate cost is spread across applications and libraries, especially when different pieces of code obscure the same underlying computation. 
We observe that seemingly unrelated code regions often perform the same computations, creating specialization opportunities that require semantic reasoning. The rise of agentic AI makes it possible to automate this semantic reasoning.
However, conventional AI agents 
cannot efficiently and reliably analyze large scale datacenter software. 

To unlock this opportunity, we present \emph{\system{}}, an agentic framework that automatically discovers hardware specialization opportunities by reasoning about what the executed code computes.  
First, \system{} extracts hot control-flow regions from execution traces across workloads and organizes them into compact analysis packages that make large workloads tractable for agents. 
Second, agents describe what each region computes, merge those that perform the same computation into \emph{computational families}, and check their semantic correctness in a loop.
Third, to verify the discovered family, \system{} runs every piece of code in a family against 
a set of reference inputs and outputs  
on recorded executions. 
Finally, for selected families, \system{} proposes hardware, e.g., ISA extensions, near-core units, or  detached accelerators,
and human experts decide which proposals warrant further development. 

We evaluate \system{} on 20 workloads across six datacenter workload categories. 
From the discovered families, 
we simulate 30 of them in  
a cycle-accurate architectural simulator and  
implement them in RTL. 
These families in hardware speed up the workloads by 12.6\% on average and by up to 80.2\%.
\end{abstract}

\maketitle

\section{Introduction}
\label{sec:introduction}

Hardware specialization can improve datacenter performance, energy, and cost efficiency~\cite{ranganathan21warehouse, abali20data, sriraman20accelerometer}. 
Specialization can take several forms, from new instructions and dedicated execution units within a CPU to tightly coupled or detached accelerators~\cite{zhang19leveraging, kanev17mallacc, karandikar21hardware, stojkovic26accelflow}.
By tailoring hardware to recurring computations, designers can avoid the overheads of executing those computations on general-purpose processors. 
At datacenter scale, even modest application-level improvements can translate into substantial savings.

However, deciding which computations warrant specialized hardware is difficult because datacenters run diverse services built from large, continuously evolving software stacks~\cite{kanev15profiling, ayers19asmdb, gonzalez23profiling}.
Their execution spans application logic, runtime systems, and many layers of libraries.
Across services, cycles spread over many functions, each contributing only part of the total cost.
Hence, a locally expensive function does not necessarily provide a useful specialization target.
Hardware vendors must consider
whether a target's aggregate benefits justify the design, integration, and deployment costs.

Successful specializations often begin with clearly identified targets.
ML provides good examples: GPUs and custom accelerators such as Google's TPU~\cite{tpu}, Meta's MTIA~\cite{krishnamurthy26mtia}, and Microsoft's Maia accelerate widely used AI/ML kernels~\cite{guthrie26maia}.
Beyond ML, researchers have identified a \emph{datacenter tax}: recurring infrastructure work, such as memory allocation, serialization, and compression, that supports many services~\cite{kanev15profiling, sriraman20accelerometer}.
Researchers proposed numerous accelerators to address these overheads~\cite{kanev17mallacc, karandikar21hardware, abali20data, pourhabibi20optimus}.
Even commercial platforms, such as Intel's Sapphire Rapids~\cite{intel22technical} and Microsoft's Cobalt 200~\cite{bilgin25announcing}, now incorporate related capabilities through integrated on-chip accelerators.
Specialization also appears within processor instruction sets, including AVX for vector operations and AMX for matrix operations~\cite{intel26architectures}.

These known targets are only the tip of the iceberg.
Finer-grained computations within and across functions and services offer additional
specialization opportunities. For example, a small
character-classification unit could accelerate a shared
step across multiple serialization and parsing functions.
However, the same computation can appear in different
forms. One implementation may compare characters
explicitly, while another uses a lookup table.
Therefore, identifying shared computations requires reasoning
about what the code computes across different implementations.
This is difficult to perform manually at datacenter scale.

Prior work has automated ISA and accelerator
discovery by identifying hot code regions or recurring
structural patterns~\cite{atasu03automatic, venkatesh10c-cores, lysecky06warp, venkatesh11qscores, trilla21novia, rezunov25cidre, xiao26isamore}. However, these approaches miss shared computations when their algorithms
differ. LLM agents offer a way to connect these
implementations via semantic reasoning. They can
interpret new code, investigate relationships
between implementations, and revise their conclusions
using analysis tools~\cite{yang24sweagent, trae25trae}. However, applying them
directly to datacenter software introduces challenges
in both scale and reliability. Large software stacks
exceed their analysis capacity, and their conclusions
can incorrectly group code with different behavior.
Hence, reliable automation requires a compact
representation of executed code and independent
correctness checks on the agents' semantic claims.

To address these challenges, we present \system{}, an
agentic framework that discovers shared computations
across workloads and translates them into hardware specialization candidates. 
\system{} makes semantic analysis scalable through compact representations of frequently executed
code regions. 
Then, agents reason about what these code regions compute
and group implementations of the same computation into
a \emph{computational family}, even when their code structure differs.
Independent agents challenge these matches and search
for overlooked relationships. 
To check the resulting
claims, \system{} tests each piece of code against
a set of provided reference inputs and outputs 
on recorded executions.

For selected families, \system{} next proposes hardware
implementations and the software changes needed to invoke
them. These candidates include ISA extensions in the core,
near-core units beside it, and detached accelerators.
\system{} chooses among them based on the amount
of work per invocation, memory-access patterns,
and the cost of invoking the hardware. 
For example, \system{} discovers hash-table probes across
database engines and proposes a parameterized instruction
that accommodates their different table layouts.
In key-value stores, it identifies recurring hash-mixing
operations across five algorithms and proposes a near-core
unit that supports those algorithms. In AI agents,
it discovers escaped-length calculations across JSON
encoders and proposes an instruction that handles their
different character policies. Human experts then assess
these proposals to determine if the benefits across
implementations justify the hardware and software
integration costs.

We evaluate \system{} on 20 workloads across six categories: key-value stores, databases, web services,
search, video transcoding, and AI agents. For each category,
we select 20 discovered families as specialization
targets. These targets cover 59.1\% of executed instructions
on average.
With the same number of targets, grouping code by function
name or dataflow graph covers only 45.4\% and 35.5\%,
respectively. Function-level targets are also often too large
and complex for hardware.
To measure the performance benefits, we implement 30 selected
families, five per workload category, as ISA extensions
and near-core units in the cycle-accurate architectural simulator and in RTL. The simulations
show that these units speed up the workloads
by 12.6\% on average and by up to 80.2\%, while the area of an average unit is only 
0.37\% of a core. 
Finally, on five \emph{production services} from a  hyperscaler fleet, \system{} 
covers the most instructions with only 20 targets, and finds previously unknown hardware specialization opportunities
that span services.

This paper makes the following contributions:
\begin{itemize}[leftmargin=*,topsep=2pt,itemsep=1pt]
    \item \system{}, the first agentic framework to automate
    the discovery of hardware specialization opportunities
    across datacenter workloads through semantic reasoning.

    \item Case studies of new instructions and hardware units
    discovered by \system{} on diverse open-source and production 
    workloads.

    \item Evaluation of 30 discovered families through
    RTL implementations and cycle-accurate architectural
    simulation.
\end{itemize}

\section{Background and Motivation}
\label{sec:backgroundmotivation}

\noindent \textbf{Hardware specialization in datacenters.} Datacenters increasingly rely on specialized hardware to improve performance and energy efficiency. Specialization takes several forms. At the coarsest grain, detached accelerators run entire
workloads. For example, GPUs and custom chips, such as Google's TPU, Meta's MTIA, and Microsoft's Maia, run ML
models~\cite{google26introduction, krishnamurthy26mtia,
guthrie26maia}, and video accelerators transcode videos~\cite{ranganathan21warehouse}. Closer to the cores, server processors integrate on-chip accelerators for data movement, compression, and encryption~\cite{intel22technical,
bilgin25announcing}. These accelerators target the \emph{datacenter tax}, the infrastructure work that many services share~\cite{kanev15profiling, sriraman20accelerometer}. At the finest grain, instruction sets add specialized instructions. For example, x86 provides an instruction for CRC-32C checksums, AVX for vector operations, and AMX for matrix
multiplication~\cite{intel26architectures}. Each form makes a different trade-off. A new instruction has almost no invocation cost, but it must fit in the pipeline and the area budget of the core. In contrast, a detached accelerator can process large amounts of data, but each invocation pays for data movement and synchronization. Hence, the right form depends on how much work a computation performs per invocation and how often it runs. In all cases, however, designers start from a
computation that they already know is worth specializing.

\begin{figure}[h]
\centering
\includegraphics[width=\columnwidth]{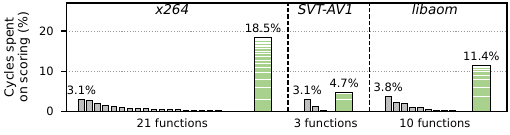}
\vspace{-7mm}
\caption{Cycles that three encoders (x264, SVT-AV1, and libaom) spend on block scoring.
}
\label{fig:satd-profile}
\end{figure}

\vspace{1pt}
\noindent \textbf{Hardware specialization opportunities extend beyond individual hotspots.}
Profilers attribute cycles to functions, and designers typically
look for the hottest ones. However, the same fine-grained computation can
appear in many functions and workloads, where each piece is too
small to attract attention on its own. For example, video
encoders score every candidate block during motion search by
summing the absolute differences between the block and its
prediction. Our measurements show that this scoring takes 18.5\%,
4.7\%, and 11.4\% of the cycles in x264~\cite{videolan26x264},
SVT-AV1~\cite{aomedia26svtav1}, and libaom~\cite{aomedia26av1},
respectively. However, as Figure~\ref{fig:satd-profile} shows,
the three encoders spread it over 34 functions: 21 in x264, 3 in
SVT-AV1, and 10 in libaom. Each gray bar is one of these
functions, and none of them takes more than 3.8\% of the cycles.
Only when the pieces are stacked together, as in the green bars,
does the scoring stand out. The reason is that each encoder implements the scoring
separately for every block size, number of reference blocks, and sample width.
Hence, a function-level profile does not reveal this
opportunity, although one hardware unit could accelerate the
scoring in all three encoders. Finding it requires gathering every
piece of code that performs the computation, across functions and
workloads.

\begin{figure}[h]
\centering
\includegraphics[width=\columnwidth]{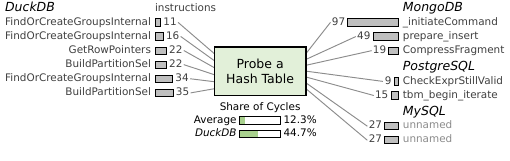}
\caption{Thirteen pieces of code in four database engines perform one computation: probing a hash table. Each piece is named by the function that contains it, and its bar shows its number of instructions.}
\label{fig:probe-fanin} 
\end{figure}

\vspace{1pt}
\noindent \textbf{Common computations can have different implementations.}
Gathering these pieces is hard, because developers implement the
same computation in many ways. For example, database engines probe
hash tables to find the slot that holds a key, or a free slot for
a new key. Figure~\ref{fig:probe-fanin} shows the pieces of code
that perform this computation in DuckDB, MongoDB, PostgreSQL, and
MySQL.
The four engines
implement the probe in 13 pieces of code, which range from 9 to
97 instructions. Even within one engine, the probe appears in
several functions: DuckDB has six pieces in three functions.
Further, the function names rarely reveal the computation. For
example, MongoDB probes a hash table inside
\texttt{CompressFragment}, a compression function, and
\texttt{\_initiateCommand}, while MySQL's functions carry no names
at all. 
Together, these pieces take 12.3\% of the cycles of the four engines
on average, and 44.7\% in DuckDB. Hence, if we recognize that
these pieces perform the same computation, a single hardware
design, such as an instruction parameterized by the table layout,
can accelerate all of them. 
Such recognition requires
reasoning about what the code computes.

\vspace{1pt}
\noindent \textbf{Agentic AI can automate the discovery of specialization opportunities.}
Agentic AI systems use LLMs to perform multi-step tasks
autonomously. An agent interleaves LLM reasoning with tool calls. 
It decides on an action, runs a tool, such as a shell command or a
program analysis, observes the result, and uses it to decide its
next action. This loop allows agents to work on tasks that do not
fit in a single prompt. Today, agents resolve issues in real
software repositories~\cite{jimenez24swebench, mini26minimal,
trae25trae}. They also increasingly design and optimize computer
systems, from algorithms and compute kernels to cache policies and
accelerators~\cite{novikov25alphaevolve, liao26kernelevolve,
gupta26archagent, nallathambi26a3d}. These capabilities match the
needs of specialization discovery. An agent can read unfamiliar
code, including the instructions of a binary, and describe what
it computes, regardless of its names or structure. It can then
search other workloads for code that performs the same
computation, and test its hypotheses with analysis tools.
However, agents are not reliable on their own. Their context
cannot hold the code of a datacenter software stack, so they see
only part of it at a time. Further, their conclusions are not
deterministic and can be wrong: an agent can group code whose
behavior differs, or miss related code. Hence, an agentic
discovery framework must give agents a compact view of the
executed code, and check their conclusions independently.

\section{\system{}: Agentic Discovery of Hardware Specialization Opportunities}
\label{sec:design}

\begin{figure}[t]
\vspace{0mm}
\centering
\includegraphics[width=\columnwidth]{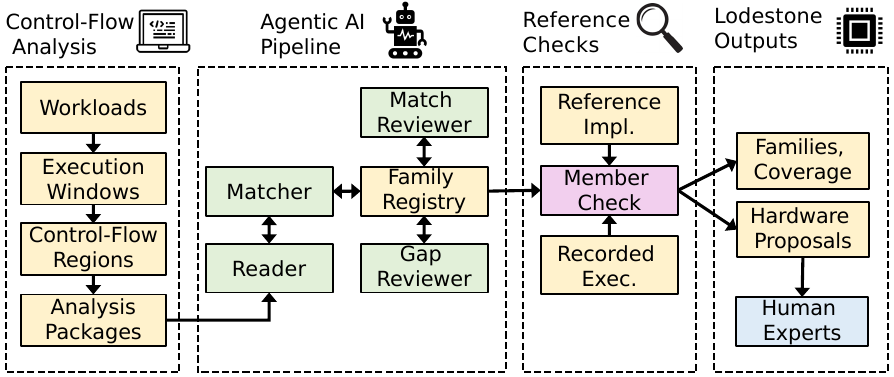} 
\vspace{-4mm}
\caption{\system{} overview: control-flow analysis extracts hot code into analysis packages, a pipeline of agents merges code that performs the same computation into families, reference checks test every member on recorded executions, and human experts assess the resulting hardware proposals.}
\label{fig:overview}
\vspace{-4mm}
\end{figure}

\system{} takes a set of workloads and discovers the computations
that they share. Its output is a set of \emph{families}. A family
consists of one computation and its \emph{members}, the pieces of
code in the workloads that perform it.  
Figure~\ref{fig:overview}
shows the four stages of \system{}. First, control-flow analysis
records execution windows of each workload, and extracts its hot
code into compact analysis packages
(Section~\ref{sec:design-regions}). Second, a pipeline of agents
reads these packages and merges the pieces of code that perform
the same computation into families
(Section~\ref{sec:design-pipeline}). Third, \system{} checks every
member against its family's reference implementation on recorded
executions (Section~\ref{sec:design-verify}). Finally, it reports
each family with its coverage and a hardware proposal, which
human experts then assess (Section~\ref{sec:design-outputs}).
These stages address the two requirements of
Section~\ref{sec:backgroundmotivation}. Control-flow analysis gives
the agents a compact view of the executed code, while the
reviewer agents and the reference checks verify the conclusions
of the agents independently.

\subsection{Control-Flow Analysis}
\label{sec:design-regions}

Datacenter binaries contain millions of instructions, but the
workloads spend most of their time in a small part of them.
Hence, \system{} first narrows the code to the hot paths,  
and packs these paths into packages
that fit in the context of an agent.

\vspace{1pt}
\noindent \textbf{Execution Sampling.} \system{} observes each
workload through \emph{execution windows}. A window is a short
sequence of user-level instructions, recorded from one worker
thread at a random time during the steady state of the workload.
\system{} records many windows per workload, each at a different
random time, and analyzes them together. Hence, the instructions
in the windows reflect where the workload spends its execution,
rather than a single phase of it. Section~\ref{sec:methodology}
gives the number and length of the windows.

\vspace{1pt}
\noindent \textbf{Control-Flow Regions.} \system{} recovers the
basic blocks and control-flow graph of each executed binary,
and maps each recorded instruction to its basic block. It then
divides the executed code into \emph{control-flow regions}. A
region is an ordered path of basic blocks that one call of a
function executes, following the edges of the recovered graph.
Hence, a region can capture a computation smaller than a
function. \system{} counts the recorded instructions that each
region covers, and keeps the hottest regions of each length for
every workload. Regions of different lengths capture computations
of different sizes: a short region can capture the
body of a hot loop, while a long one can capture the path that
leads into the loop.

\begin{figure}[t]
\centering
\includegraphics[width=\columnwidth]{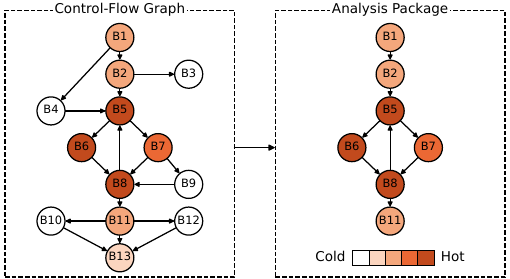}
\caption{Control-flow graph of a function and the analysis package that keeps its hot control flow.}
\label{fig:packages}
\vspace{-2mm}
\end{figure}

\vspace{1pt}
\noindent \textbf{Analysis Packages.} The hot regions of a
function overlap, as they share most of their basic blocks. If
agents analyzed each region separately, they would read the same
blocks many times. Instead, \system{} groups overlapping regions
into \emph{analysis packages}. A package stores each basic block
once, together with its instructions, control-flow edges, and
execution counts, and lists the paths that its regions follow
through these blocks. Figure~\ref{fig:packages} shows an example.
The control-flow graph of the function has 13 basic blocks, but
most of its execution stays in the loop of blocks B5--B8. The
package keeps only this loop and the hot paths into it (B1 and
B2) and out of it (B11), and drops the blocks that rarely
execute.
Hence, a package keeps the input
of an agent small, while preserving every instruction and path
that the agent needs to understand the code. Finally, \system{}
indexes all packages by shared basic blocks and by similar
instruction sequences. As a result, an agent can locate related
code across workloads without reading every package.

\subsection{Agentic AI Pipeline}
\label{sec:design-pipeline}

The agentic AI pipeline is the core of \system{}. It reads the
analysis packages and discovers which pieces of code, across
functions and services, perform the same computation. This step
needs semantic reasoning, as the pieces of one computation
can differ in names, algorithms, control flows, and data
layouts. Agents can
perform such reasoning, but a single agent cannot perform it
reliably at scale. The packages of one workload category do not fit
in its context. Further, an agent that interprets a piece of code
early tends to fit later code to this interpretation, and an
agent that both proposes and approves a match has no one to catch
its mistakes. Hence, \system{} organizes the pipeline as a review
process among four agent \emph{roles}, i.e., four types of
agents, each with its own task and its own context, shown in
the middle of Figure~\ref{fig:overview}. Readers describe the
code, the Matcher proposes families, and two reviewers check the Matcher's
conclusions: the Match Reviewer looks for false
matches, and the Gap Reviewer for missed ones. The roles
coordinate through a shared family registry. \system{} runs one
pipeline per workload category
and 
processes 
all workloads in
the category together.

\vspace{1pt}
\noindent \textbf{Family Registry.} The agents of different roles
do not share their contexts. Instead, they share the state of the
pipeline through a \emph{family registry}, which is stored
outside the agents. For each family, i.e., a computation and the
pieces of code that perform it, the registry records the
computation, its inputs, and effects. It also lists the
members of the family with their exact locations in the recorded
windows. The registry also holds an
investigation plan. The plan has one \emph{task} for every pair
of a package and a target workload in the category. A task asks
whether the target workload contains code that performs the
computation of the package. Hence, the plan compares every
package with every workload, instead of only the pairs that an
agent happens to consider. A task remains open until its
investigation has a reviewed outcome. Each role takes its work
from the registry and writes its results back. As a result, an
agent needs only the context of its current task, the pipeline
scales to thousands of regions, and it can resume after any step.

\vspace{1pt}
\noindent \textbf{Reader.} Readers describe the instructions of
each package in notes that later agents can search. A Reader
examines every basic block and path in its packages, and writes a
note for every instruction position, i.e., for each instruction
of each basic block. The notes describe what the code computes,
its inputs and effects, the conditions under which each path
runs, and what remains unknown. Each note references the
instructions that it is based on, so that other agents can check
it against the code. \system{} indexes the notes together with the
packages. Hence, a later agent can search for code by what it
computes, and can start from a description of the code instead of its raw
instructions.
\vspace{0mm}

\vspace{1pt}
\noindent \textbf{Matcher.} It works through the open
tasks of the plan. For each task, it searches the indexed packages
and notes of the target workload for code that performs the
computation of the package. Since notes can be incomplete or
wrong, it reopens the original instructions of every
candidate before it draws a conclusion. It then proposes a new
family, or new members of an existing family. A family proposal
states the inputs, computation, and effects of the family, lists
the members that perform it, and describes how they differ.
Members may use different algorithms, control flows, or data
layouts. When members differ in a parameter of the computation,
such as the hash function or the table layout, each value of the
parameter becomes a separate \emph{mode} of the family. For
example, for a DuckDB package that probes a hash table, the
Matcher searches MongoDB and finds the probe in
\texttt{prepare\_insert}. If the two
probes use different table layouts, they become two modes of the
same family. Each investigation ends with a match, no match, or an
open question, when the available evidence cannot settle the task.

\vspace{1pt}
\noindent \textbf{Match Reviewer.} It guards
against false matches. It examines every match that the Matcher
proposes. To avoid adopting the interpretation of the Matcher, it
first opens the code of each proposed member and writes its own
account of what the code computes, and only then reads the
proposal. It checks that each member performs the claimed
computation, and that all members share it. It then either
approves the family, or returns specific objections that the
Matcher must answer with evidence or a revision.

\vspace{1pt}
\noindent \textbf{Gap Reviewer.} It guards against
missed matches and examines investigations that end without
a match. It works out the computation of the package on its own,
and searches all packages of the target workload for code that
performs it, including the packages that the Matcher did not
examine. When it finds such code, the task returns to the
Matcher, and any new match goes to the Match Reviewer. For an
open question, the Gap Reviewer checks if the missing
evidence is available. If so, it returns the task to the Matcher
for further investigation. Otherwise, it accepts the outcome.

\subsection{Reference Checks}
\label{sec:design-verify}

Reviewers reduce the mistakes of the agents, but they cannot
eliminate them. Hence, before \system{} reports a family, it
checks every member of the family on real executions.

\vspace{1pt}
\noindent \textbf{Reference Implementations.} For each family, a
\emph{reference implementation} is an executable program that
specifies the computation of the family in every mode.
We write it from the registry entry of the family and
representative code of its members. We do not look at any
recorded value, and we freeze the reference before any check
runs. Hence, the reference captures the claimed computation,
and not the behavior of the members on the checked
executions.

\vspace{1pt}
\noindent \textbf{Member Checks.} To check a member, \system{}
re-runs its workload, and records each execution of the code of
the member, together with the values that the code reads and
writes. For each recorded execution, \system{} gives the values
that the member reads to the reference, and checks that the
reference produces the values that the member writes. To do so,
\system{} maps each input and result of the reference to the
register or memory location where the member keeps it. The member
passes if it matches the reference on every
recorded execution. After a failure, we can fix the reference or
this mapping, but we check the fix only on new executions.
\system{} removes the members that still fail from the family.

\subsection{\system{} Outputs}
\label{sec:design-outputs}

\system{} reports each family with its reference implementation,
 modes, members, and coverage, i.e., the share of 
instructions and cycles per workload that its members cover.
Since a family combines members from many functions and
workloads, its coverage shows the aggregate cost that hardware for
the family could address. For each family, \system{} also prepares a hardware proposal. The proposal
specifies the operations of the hardware, the parameters that
distinguish the modes of the family, such as constants or table
layouts, and the point at which each member invokes the hardware.
Depending on the work per invocation, the memory behavior, and the
invocation cost of the family, the hardware can be an ISA
extension, a near-core unit, a detached accelerator, or a
combination of these. The proposal also lists the software
changes that each member needs. As we will explain in Section~\ref{sec:usecases}, human experts then decide which
proposals justify their design and integration costs.

\subsection{Programmability}
\label{sec:design-programmability}

\system{}'s hardware is simple to program. An ISA extension
is a new instruction. For example,
\texttt{hprobe rd, rb, rh, rt, mode} probes the table at
\texttt{rb} from hash \texttt{rh}, and returns in \texttt{rd}
the slot of tag \texttt{rt}, while \texttt{mode} encodes the table
layout. A near-core unit is invoked with an ATX-~\cite{gerogiannis26atx} or RoCC-like
instruction~\cite{asanovic16rocket,
karandikar21hardware, karandikar23cdpu}. For example,
\texttt{hmix rd, ra, rl, mode} returns in \texttt{rd} the hash
of the \texttt{rl}-byte key at \texttt{ra}. A detached
accelerator receives descriptors, like Intel's DSA~\cite{yuan24intel, kuper24quantitative}.
\system{} records the code addresses, mode, and input and
result locations of each member. Hence,
(i) a programmer can use these as hints, or (ii)  
a compiler or binary
optimizer~\cite{panchenko19bolt} can replace each member
automatically with code that moves the inputs into place,
issues the instruction, and moves the result back.

\section{Methodology}
\label{sec:methodology}

\begin{table}[t]
\centering
\footnotesize
\caption{Evaluated workloads and their configurations.}
\label{tab:workloads}
\begin{tabularx}{\columnwidth}{@{}
    p{0.20\columnwidth}
    p{0.18\columnwidth}
    X@{}}
\toprule
Category & Workload & Description \\
\midrule

Key-Value Stores & Memcached~\cite{memcached26memcached} & CloudSuite data-caching load~\cite{ferdman12clearing, palit16demystifying} on a 10\,GB Twitter dataset \\
  & Redis~\cite{redis26documentation} & YCSB workload A~\cite{cooper10benchmarking} over 10M records \\
 & RocksDB~\cite{dong21evolution} & db\_bench~\cite{rocksdb26benchmarking} with the benchmark-script options, 50\% writes, and 20M keys \\
 & TaoBench~\cite{su25dcperf} & Tao-patched Memcached from DCPerf~\cite{su25dcperf}; load at a steady 90\% hit rate \\
\midrule
Databases & DuckDB~\cite{raasveldt26duckdb} & 22 TPC-H~\cite{tpc26tpch} queries at scale factor 1000 \\
 & PostgreSQL~\cite{postgresql26documentation} & TPC-C~\cite{tpc26tpcc}: 100 warehouses in BenchBase~\cite{difallah13oltpbench} \\
 & MySQL~\cite{oracle26mysql} & TPC-C: 100 warehouses in BenchBase \\
 & MongoDB~\cite{mongodb26manual} & YCSB workload A over 10M records \\
\midrule
Web Services & Nginx~\cite{nginx26documentation} & OpenResty~\cite{openresty26openresty} front end of the DeathStarBench social network~\cite{gan19open}, driven by wrk2~\cite{tene26wrk2} \\
 & MediaWiki~\cite{mediawiki26manual} & Runs on PHP-FPM~\cite{php26fastcgi} and serves the DCPerf oss-performance load~\cite{su25dcperf} \\
 & Envoy~\cite{envoy26documentation} & Nighthawk~\cite{envoy26nighthawk} load of 20K RPS \\
 & Django~\cite{django26documentation} & Runs on uWSGI~\cite{uwsgi26project} and CPython~\cite{python26documentation} and serves DCPerf DjangoBench~\cite{su25dcperf} \\
\midrule
Search & HDSearch & $\mu$Suite~\cite{sriraman18usuite} image similarity search over 2.8\,GB of feature vectors \\
 & SetAlgebra & $\mu$Suite posting-list intersection over 11.4\,GB of posting lists \\
 & Recommend & $\mu$Suite collaborative filtering over 26M ratings \\
\midrule
Video & x264~\cite{videolan26x264} & FFmpeg~\cite{ffmpeg26documentation} with a 4K 10-bit, 1,199-frame clip \\
Transcoding & SVT-AV1~\cite{aomedia26svtav1} & Same clip and settings \\
 & libaom~\cite{aomedia26av1} & Same clip and settings \\
\midrule
AI Agents & MiniSWE~\cite{mini26minimal} & mini-swe-agent on CPython solves 20 SWE-bench Verified~\cite{jimenez24swebench, chowdhury24introducing} tasks \\
 & Trae~\cite{trae25trae} & trae-agent on CPython; same 20 tasks \\

\bottomrule
\end{tabularx}
\end{table}

\vspace{1pt}
\noindent \textbf{Setup.} We study 20 workloads in six categories:
key-value stores, databases, web services, search, video transcoding, and AI agents.
Table~\ref{tab:workloads} lists the workloads and their configurations.
All workloads run on
Intel Xeon Platinum 8581C
processors and Ubuntu 22.04.

\vspace{1pt}
\noindent \textbf{Execution Windows.} We record execution windows with DynamoRIO's
drmemtrace~\cite{bruening03dynamorio}. For each workload, we record 20 \emph{discovery} windows and
20 \emph{held-out} windows in separate runs. Each window has 50M user-level instructions.
\system{} sees only the discovery windows. We use the held-out windows to check whether the families
also cover new executions, and to measure performance. To measure cycles, we also run each workload
natively, without tracing, under Linux perf~\cite{linux26perf}. 
The share of perf samples that fall in a piece of code estimates its share of the user-level cycles. 

\vspace{1pt}
\noindent \textbf{\system{} Configuration.} We run \system{} once per category, on the discovery windows
of all its workloads together. All agents use OpenAI GPT-6 Astra~\cite{openai26gpt6} at high reasoning effort.
Control-flow regions contain 1 to 32 basic blocks, and for each workload, \system{} keeps the 100 hottest
regions of each length. The member checks test each member on at least 30 of its recorded executions.

\vspace{1pt}
\noindent \textbf{Baselines.} We compare \system{} with six methods that group the hot code into \emph{targets},
i.e., candidates for hardware. \emph{Region} keeps each hot control-flow region as its own target,
as before the agents merge them. \emph{Func} groups the regions by function, and \emph{FuncName} groups
them by function name across workloads. \emph{DFG-Exact} groups the regions that have the same dataflow
graph with exact operand labels, and \emph{DFG-Class} relaxes these labels to opcode classes.
Finally, \emph{Tax} groups the regions into 11 datacenter tax categories based on~\cite{kanev15profiling, sriraman20accelerometer}.
All methods start from the same hot regions of the discovery windows,
and differ only in how they group them. Hence, the comparison isolates the effect of the grouping.

\vspace{1pt}
\noindent \textbf{Coverage.} We compare the methods by the coverage of their targets. Each method gets a budget of
N targets (from 1 to 100). Every method picks its targets with the same greedy selector, which repeatedly adds
the target that covers the most instructions not yet covered in the discovery windows. We report coverage in two ways:
the share of the executed instructions in the discovery windows that the chosen targets cover, and their share of the
user-level cycles.

\begin{table}[h]
\centering
\footnotesize
\caption{Configuration of the simulated core.}
\label{tab:sim}
\begin{tabular}{@{}ll@{}}
\toprule
Frequency & 3.1\,GHz \\
Width & 6-wide decode/rename/dispatch, 8-wide retire \\
Out-of-order & 512-entry ROB, 192/114 load/store queue \\
Front end & TAGE predictor, 12K-entry BTB, 4K $\mu$op cache \\
L1I & 32\,KB, 8-way, 3-cycle round-trip (RT) latency \\
L1D & 48\,KB, 12-way, 5-cycle RT latency \\
L2 & 2\,MB, 16-way, 16-cycle RT latency \\
LLC & 5\,MB per core, 20-way, 96-cycle RT latency \\
Memory & DDR5-5600 timing, 2 channels (Ramulator~\cite{kim16ramulator}) \\
\bottomrule
\end{tabular}
\end{table}

\begin{table*}[t]
\centering
\footnotesize
\setlength{\tabcolsep}{3pt}
\caption{The five representative families of each category, with the number of workloads, \emph{FuncName} groups, and \emph{DFG-Class} groups that their code spans, and the largest speedup of their code. Families marked with $\dagger$ are discussed in Sections~\ref{sec:case-kv}--\ref{sec:case-agents}.}
\vspace{-4mm}
\label{tab:families}
\begin{tabular}{@{}llllrrrrrr@{}}
\toprule
Category & Family & Description & Form & Members & Workloads & \emph{FuncName} & \emph{DFG-Class} & Speedup & Area ($\mu$m$^2$) \\
\midrule
Key-Value Stores & Hash Mixing$^\dagger$ & Mixes key bits into a hash & Near-core & 8 & 4 & 6 & 7 & 5.50$\times$ & 5,658 \\
 & Object Allocation & Takes and returns free objects & ISA ext. & 29 & 4 & 25 & 29 & 1.99$\times$ & 2,690 \\
 & Cursor Advance & Moves a cursor to the next record & ISA ext. & 11 & 4 & 11 & 11 & 1.22$\times$ & 5,203 \\
 & Keyed Lookup & Finds a key in a hash table & ISA ext. & 11 & 4 & 9 & 11 & 1.22$\times$ & 9,058 \\
 & Frame Decoding & Decodes a field from a byte stream & ISA ext. & 6 & 4 & 5 & 6 & 3.51$\times$ & 9,587 \\
\midrule
Databases & Hash Probe$^\dagger$ & Probes a hash table for a match & ISA ext. & 13 & 4 & 10 & 16 & 6.23$\times$ & 3,415 \\
 & Value Hashing & Hashes a value or byte string & ISA ext. & 13 & 4 & 10 & 14 & 3.52$\times$ & 9,623 \\
 & Byte Scan \& Filter & Selects values that match a predicate & ISA ext. & 23 & 4 & 13 & 21 & 2.32$\times$ & 8,838 \\
 & Bit Unpacking & Unpacks bit-packed integers & Near-core & 19 & 4 & 18 & 22 & 1.64$\times$ & 13,936 \\
 & Aggregate Update & Folds an input into an aggregate & Near-core & 10 & 4 & 9 & 10 & 2.22$\times$ & 41,802 \\
\midrule
Web Services & Hash Slot Location$^\dagger$ & Finds the slot of a key & ISA ext. & 17 & 4 & 15 & 16 & 2.10$\times$ & 1,920 \\
 & Free-List Push/Pop & Pushes or pops a list head & ISA ext. & 15 & 4 & 12 & 15 & 1.56$\times$ & 2,401 \\
 & Affine Accumulation & Accumulates values with multiply and add & Near-core & 12 & 4 & 7 & 11 & 3.95$\times$ & 6,352 \\
 & Counter Decrement & Decrements a counter and tests for zero & ISA ext. & 11 & 4 & 7 & 9 & 2.56$\times$ & 368 \\
 & Hash Entry Read & Reads the value of a found key & ISA ext. & 20 & 3 & 13 & 19 & 2.76$\times$ & 2,018 \\
\midrule
Search & Tree Descent$^\dagger$ & Walks down a search tree & Near-core & 12 & 2 & 3 & 5 & 2.67$\times$ & 44,077 \\
 & Array Packing & Copies and packs array data & Near-core & 19 & 3 & 12 & 19 & 1.21$\times$ & 16,898 \\
 & Field Encoding & Encodes and decodes stored fields & ISA ext. & 34 & 3 & 14 & 37 & 1.38$\times$ & 1,788 \\
 & Malloc Metadata & Updates allocator metadata & ISA ext. & 7 & 3 & 3 & 7 & 1.31$\times$ & 637 \\
 & List Splice & Inserts and removes list nodes & ISA ext. & 14 & 3 & 10 & 15 & 1.04$\times$ & 2,397 \\
\midrule
Video Transcoding & SATD Scoring$^\dagger$ & Sums absolute transformed differences & ISA ext. & 119 & 3 & 34 & 213 & 3.51$\times$ & 48,001 \\
 & Symbol Rate Cost & Estimates the bits of coded coefficients & Near-core & 369 & 3 & 7 & 799 & 2.41$\times$ & 92,284 \\
 & Sample Filtering & Filters samples with weighted sums & Near-core & 159 & 3 & 22 & 291 & 3.60$\times$ & 265,903 \\
 & Integer Transform & Applies integer DCT and Hadamard transforms & Near-core & 32 & 3 & 18 & 30 & 2.73$\times$ & 172,946 \\
 & Residual Statistics & Sums residuals and their squares & ISA ext. & 157 & 3 & 19 & 316 & 2.32$\times$ & 116,734 \\
\midrule
AI Agents & Escape Length$^\dagger$ & Counts a JSON string's escaped length & ISA ext. & 3 & 2 & 2 & 14 & 4.46$\times$ & 1,104 \\
 & Escape Copy$^\dagger$ & Copies a string with JSON escaping & Near-core & 5 & 2 & 4 & 15 & 6.11$\times$ & 3,080 \\
 & Dict Probe & Looks up a key in a Python dict & ISA ext. & 11 & 2 & 3 & 11 & 1.41$\times$ & 1,701 \\
 & Node Unlink & Removes a node from a linked list & ISA ext. & 4 & 2 & 2 & 4 & 1.19$\times$ & 1,262 \\
 & Bucketed Free & Returns a block to a free list & ISA ext. & 3 & 2 & 2 & 3 & 1.43$\times$ & 1,166 \\
\bottomrule
\end{tabular}
\end{table*}

\vspace{1pt}
\noindent \textbf{RTL Design.} We design and implement the hardware proposals in RTL and synthesize them with
Yosys~\cite{wolf26yosys} using the ASAP7 7~nm library~\cite{clark16asap7, vashishtha17asap7}. 
We compare the
areas with an Emerald Rapids core, which we estimate at 8~mm$^2$ with its 2\,MB L2~\cite{locuza22die}.
We also estimate the area of one more ALU and of 128 more ROB entries with McPAT~\cite{li09mcpat, li11cactip}, scaled to the core,
and the area of 1\,MB more L2 from the measured L2 density~\cite{locuza22die}.

\vspace{1pt}
\noindent \textbf{Performance Simulation.} We measure the end-to-end application performance with
Scarab~\cite{scarab, oh24udp}, configured after an Emerald Rapids core (Table~\ref{tab:sim}).
Scarab replays instructions in the recorded window and counts its cycles. We simulate the held-out windows,
warm up each window for 10M instructions, and measure the rest. We model the units of one category together.
To model a unit, we replace every execution of the code of a family in the window with operations on the unit. 
These operations keep every memory access of the code, which Scarab simulates through the cache hierarchy, and accelerate only the computation according to the unit’s RTL design.
For near-core units, we model the overheads of exchanging the operands and results with the core. 
End-to-end speedup is the percentage by which the
total user-level cycles on the unmodified core exceed the total cycles with the units. 
We simulate the core with one more ALU, 128 more ROB entries, or 1\,MB more L2.

\section{Discovered Specialization Opportunities}
\label{sec:usecases}

In this section, we present the specialization opportunities that \system{} discovers in the 20 workloads.
We first summarize the families that \system{} finds across the six categories. We then select 30 of these families,
five per category, and implement them in hardware. Finally, we present seven of them
in detail, two in AI agents and one in each other category (Sections~\ref{sec:case-kv}--\ref{sec:case-agents}).

\vspace{1pt}
\noindent \textbf{Overview.}
For each workload, \system{} starts from the 100 hottest control-flow regions of each length, from 1 to 32 basic blocks,
in the discovery windows (Section~\ref{sec:methodology}). \system{}'s agentic AI pipeline merges the regions that perform the
same computation into families, and the reference checks test every member of each family.
Table~\ref{tab:reach} summarizes the result. Across the six categories, \system{} finds 254 families that merge
several hot regions, from 23 in video transcoding to 97 in databases. These families have 2,506 members,
which \system{} checks on 1,365,644 recorded executions. The members of a family often come from different
functions and have different dataflow graphs.
To quantify this, we analyze how the code of each family spreads over function names and dataflow graphs.
For each family, we find the function name that holds most of its instructions.
On average, 51.7\% of the instructions of a family lie outside this function name.
Similarly, 58.2\% of them lie outside the dataflow graph, with opcode classes as labels, that holds most of the family.
Table~\ref{tab:reach} reports these shares, together with the shares of cycles,
in the \emph{FuncName} and \emph{DFG-Class} columns. Hence, most of the code of a typical family lies
outside any single function name or dataflow graph.
As a result, a designer who examines one function or one dataflow graph at a time sees, on average, less than
half of the code of a family. In contrast, \system{} presents all the code of a family as one candidate for hardware.

\begin{table}[h]
\centering
\footnotesize
\caption{Families that merge several hot regions, with their members and checked executions. The last two columns show the average share (\%) of the instructions / cycles of a family outside its largest function name and dataflow graph.} 
\vspace{-4mm}
\label{tab:reach}
\setlength{\tabcolsep}{2.5pt}
\begin{tabular}{@{}lrrrcc@{}}
\toprule
 & & & & \multicolumn{2}{c}{Outside largest} \\
\cmidrule(lr){5-6}
Category & Fam. & Members & Executions & \emph{FuncName} & \emph{DFG-Class} \\
\midrule
Key-Value Stores & 26 & 207 & 279,366 & 47.7 / 49.5 & 48.9 / 53.0 \\
Databases & 97 & 272 & 44,535 & 63.5 / 60.0 & 68.6 / 59.2 \\
Web Services & 30 & 367 & 567,337 & 68.2 / 77.4 & 72.3 / 81.4 \\
Search & 27 & 242 & 338,524 & 37.1 / 45.3 & 39.1 / 49.5 \\
Video Transcoding & 23 & 1,196 & 107,659 & 66.5 / 82.6 & 73.4 / 81.5 \\
AI Agents & 51 & 222 & 28,223 & 27.5 / 17.0 & 46.9 / 36.2 \\
\midrule
All & 254 & 2,506 & 1,365,644 & 51.7 / 55.3 & 58.2 / 60.2 \\
\bottomrule
\end{tabular}
\vspace{-5mm}
\end{table}

Next, we act as the human experts who decide which hardware proposals to implement (Section~\ref{sec:design-outputs}).
In each category, we select five families that represent the category, recur across its workloads, and benefit from
specialization. Table~\ref{tab:families} lists these 30 families, with the form of hardware that \system{}
proposes for each, the number of members, and the area of our implementation. 
The rest of this section presents the seven families marked with $\dagger$ in Table~\ref{tab:families}.

\subsection{Key-Value Stores}
\label{sec:case-kv}

For key-value (KV) stores, \system{} discovers the \emph{Hash Mixing} family,
i.e., the code that hashes keys and other words in all workloads, and we
implement this family as a near-core unit. 
KV stores hash the key
on a table lookup. \system{} finds
that eight members across four workloads perform the same
computation. They mix the words of an input into a 32- or 64-bit
value with a fixed sequence of multiplies, shifts, rotations, and
XORs. The modes of the family cover five algorithms. Memcached and TaoBench
use MurmurHash3, Redis uses SipHash-1-2, in a case-sensitive and a
case-insensitive version, and a six-field fingerprint, and RocksDB
uses its own 32-bit hash and Mersenne Twister tempering. The members span six function names and seven dataflow
graphs. They take 14.2\% of the cycles in Memcached, 7.4\% in TaoBench, 3.6\% in Redis, and 0.4\% in RocksDB.

\begin{figure}[h]
\centering
\includegraphics[width=\columnwidth]{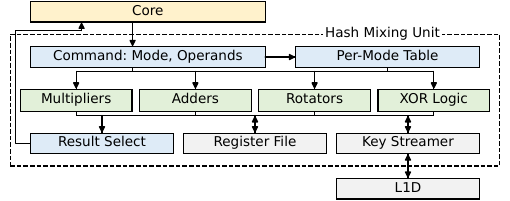}
\caption{Hash Mixing unit for key-value stores.}
\label{fig:case-kv}
\end{figure}

\vspace{1pt}
\noindent \textbf{How One Unit Serves Them.} Each hash reads a key
from memory and runs several rounds of mixing over its words. Hence,
\system{} proposes a near-core unit, which sits next to the core and
reads the key from the L1 data cache. Figure~\ref{fig:case-kv} shows
the unit that we implement. Since the five algorithms use the same
operations, the unit runs all of them on one shared datapath. To
hash a key, the core sends the unit a command with a mode, which
selects the algorithm, and the operands, such as the address and
the length of the key. A key streamer fetches the key from the L1
data cache, aligns it into 32- or 64-bit words, and masks the last
partial word.
A mixing stage then
processes the words with shared multipliers, adders, rotators, and
XOR logic, and keeps the hash state in a small register file. Each
member must produce exactly the hash values that its software and
its stored data depend on. Hence, a per-mode table supplies the
multiply constants, the rotation amounts, and the order of the
operations of each algorithm. The unit returns the hash to
the core. 
The whole unit takes 
5,658\,$\mu$m$^2$. To use the unit, each member replaces its hashing
code with a command to the unit.

\vspace{1pt}
\noindent \textbf{Benefits.} Memcached hashes the key of every
request before it accesses its table. The unit speeds up 
hashing by 5.49$\times$ and contributes 13.8 percentage points of
the 19.5\% end-to-end speedup of Memcached. The same unit also
runs the hashes of the other three workloads. 
In contrast, prior
hash-table hardware computes its own fixed hash~\cite{gope17architectural, zhang19leveraging},
and x86 offers instructions only for CRC-32C and cryptographic hashes~\cite{intel26architectures}. Neither can run these five
algorithms, so neither can serve these workloads without changing
the hash values that they depend on.

\subsection{Databases}
\label{sec:case-db}

For databases, \system{} discovers the \emph{Hash Probe} family, i.e., the
code that probes hash tables, which we
implement as an ISA extension. Databases probe hash
tables to join and aggregate rows, and to find a free slot for a
new key. The family has 13 members, the pieces of code in
Figure~\ref{fig:probe-fanin}. Each member
computes the address of a slot from a hash or an index, tests the
tag, control byte, or bit of the slot, and advances to the next
slot on a miss. The members differ in table layout 
and in how they test a slot. DuckDB probes linearly over pointers
that carry a salt of the hash, MongoDB inserts into SwissTables
that keep one control byte per slot, PostgreSQL probes backward
over status bytes, and MySQL scans bitmaps. The members' code
spans 10 function names and 16 dataflow graphs, and takes 12.3\% of the cycles of the four workloads on average and 44.7\% in DuckDB.

\begin{figure}[h]
\centering
\includegraphics[width=\columnwidth]{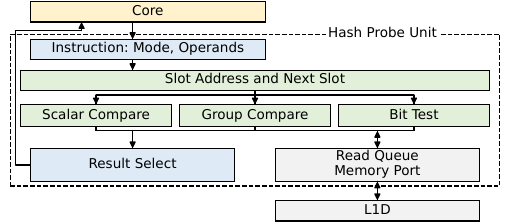}
\caption{Hash Probe unit for databases.}
\label{fig:case-db}
\end{figure}

\vspace{1pt}
\noindent \textbf{How One Unit Serves Them.} Each probe
performs little work, and the core immediately branches on its
result. Hence, \system{} proposes an ISA extension. Figure~\ref{fig:case-db} shows the unit
that executes the instruction. Its operands include the base
address of the table, the hash or index, and the tag or key to look
for. Its mode encodes the layout of the table: the size of
a slot, the position of the tag, and the direction of the probe.
First, the unit computes the address of the slot, and reads the
slot through a memory port.
Then, it tests the slot with one of three comparators, which the
mode selects. A scalar compare tests a single tag, such as the salt
of a DuckDB pointer or a PostgreSQL status byte. A group compare
tests the control bytes of a group of SwissTable slots at once. A
bit test checks the bits of a MySQL bitmap. On a miss, the unit
advances to the next slot. Finally, the result select returns the
slot to the core. All 13 members share the address and next-slot
logic and differ only in their mode, and the whole unit takes
3,415\,$\mu$m$^2$. To use the unit, each member replaces its probe
loop with the instruction.

\vspace{1pt}
\noindent \textbf{Benefits.} In DuckDB, the probes run within the
hash joins and aggregations of the TPC-H queries. The unit speeds
them up by 6.23$\times$ and contributes 67.7 percentage points of
the 80.2\% end-to-end speedup of DuckDB. In contrast, prior
database hardware probes hash indexes only for
lookups~\cite{kocberber13meet}, and prior hash-table instructions
require a table format of their own~\cite{zhang19leveraging}. The
unit also inserts into tables and counts partitions, and it works
directly on the existing tables of each engine, such as the salted
pointers of DuckDB and the control bytes of SwissTable.

\subsection{Web Services}
\label{sec:case-web}

For web services, \system{} discovers the \emph{Hash Slot Location}
family, i.e., the code that finds the slot of a key in a table, which
we implement as an ISA extension. Web services look up
keys in the tables of their language runtimes, such as Lua tables,
PHP arrays, and Python dictionaries. The family has 17 members
across four workloads. Each member selects a candidate slot from a
key, checks the key stored there, and either returns the slot or
advances to the next candidate. The members include node probes in
LuaJIT for Nginx, bucket lookups in PHP for MediaWiki, method caches
and compact dictionaries in CPython for Django, and a binary search
over a sorted byte table in Envoy. As a result, they span 15
function names and 16 dataflow graphs. In Nginx alone, eight members
in six LuaJIT functions take 11.8\% of the cycles.

\begin{figure}[h]
\centering
\includegraphics[width=\columnwidth]{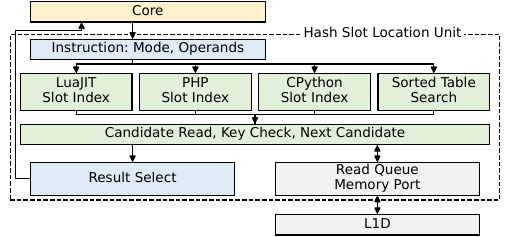}
\caption{Hash Slot Location unit for web services.}
\label{fig:case-web}
\end{figure}

\vspace{1pt}
\noindent \textbf{How One Unit Serves Them.}
Figure~\ref{fig:case-web} shows the unit that executes the
instruction. The mode of the instruction selects the table format.
For each format, a small block computes the first candidate slot of
the key: one block each for LuaJIT, PHP, and CPython, and one that
searches the sorted table of Envoy. All formats then share one
pipeline, which reads the candidate through the memory port, checks
its key, and moves to the next candidate on a miss. The
result select returns the slot to the core. The whole unit takes
1,920\,$\mu$m$^2$.

\vspace{1pt}
\noindent \textbf{Benefits.} The unit speeds up these lookups by
1.67$\times$ in Nginx.
In contrast, prior hardware for dynamic languages
serves the arrays of one PHP engine~\cite{gope17architectural}, and
prior hash-table instructions require a table format of their
own~\cite{zhang19leveraging}. The unit works on the tables that
these runtimes already use.

\subsection{Search}
\label{sec:case-search}

For search, \system{} discovers the \emph{Tree Descent} family,
i.e., the code that descends a search tree, which we implement as
a near-core unit. 
The family has 12 members across two workloads. Each
member performs one step down a red-black tree of the C++ standard
library. It reads the key and the two child pointers of a node,
compares the key with the query, and moves to the left or right
child. HDSearch runs these steps when its FLANN library looks up
the hash buckets of image features, and SetAlgebra when it collects
the document IDs of a query. The compiler inlines the lookup into
each caller, and the copies differ in key width and loop
layout. The members span three function names and five
dataflow graphs. The lookups take 22.2\% and 41.9\% of the cycles of
HDSearch and SetAlgebra, but a function-level profile
attributes these cycles to the callers.

\begin{figure}[h]
\centering
\includegraphics[width=\columnwidth]{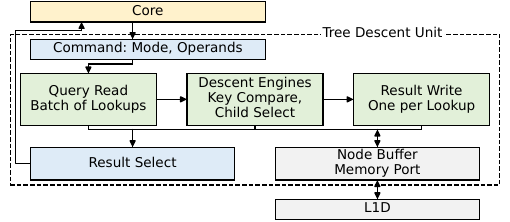}
\caption{Tree Descent unit for search.}
\label{fig:case-search}
\end{figure}

\vspace{1pt}
\noindent \textbf{How One Unit Serves Them.}
Figure~\ref{fig:case-search} shows the unit we implement. The
core sends the unit a command with the operands: the root of
the tree and a batch of queries. The unit first reads the queries of
the batch. Its descent engines then walk the tree for all of them.
At each node, an engine compares the key of the node with its query
and selects the child. The lookups of a batch share the upper levels
of the tree, so the node buffer serves these nodes to many lookups.
The unit writes one result per lookup and notifies the
core. The whole unit takes 44,077\,$\mu$m$^2$, i.e., 0.55\% of a
core.

\vspace{1pt}
\noindent
\textbf{Benefits.} It alone speeds up HDSearch and SetAlgebra by 16.1\% and 10.0\%.
Prior traversal
hardware extends GPU ray-tracing
units~\cite{ha24generalizing, barnes24extending} or implements
programmable traversal engines~\cite{lee24terminus}, and no
mainstream ISA compares a key and picks a child in one
step.

\subsection{Video Transcoding}
\label{sec:case-video}

For video transcoding, \system{} discovers the \emph{SATD Scoring}
family, i.e., the block scoring of Figure~\ref{fig:satd-profile}, which we implement as an ISA extension.
The family has 119 members
across the three workloads. Each member subtracts a predicted block
from the source block, applies a fixed linear transform, and sums
the absolute values. With the identity transform, this is the sum of
absolute differences (SAD), which all three encoders compute against
one to four reference blocks with 8- or 16-bit samples. With the
Hadamard transform, this is the sum of absolute transformed
differences (SATD), which x264 computes on blocks from 4$\times$4 to
16$\times$16. SVT-AV1 and libaom also sum the absolute values of
transform coefficients. The members' code spans 34 function names and 213 dataflow graphs.

\begin{figure}[h]
\centering
\includegraphics[width=\columnwidth]{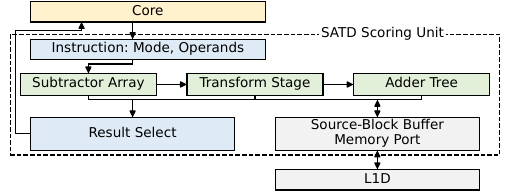}
\caption{SATD Scoring unit for video transcoding.}
\label{fig:case-video}
\end{figure}

\vspace{1pt}
\noindent \textbf{How One Unit Serves Them.}
Figure~\ref{fig:case-video} shows the unit that executes the
instruction. 
The unit runs all members on one shared datapath. 
The mode of the
instruction selects the transform, the block size, the sample width,
and the number of reference blocks. Its operands locate the source block and the predicted blocks. The unit reads the
blocks through a memory port to the L1 data cache. It keeps the
source block in a buffer, since an encoder scores many candidates
against the same source block. First, an array of subtractors
computes the differences between the source block and up to four
predicted blocks. For the coefficient sum, the prediction is zero.
Then, a transform stage applies the Hadamard transform for SATD, or
passes the differences unchanged for SAD and the coefficient sum.
Finally, an adder tree sums the absolute values, and the result
select returns the score to the core. The whole unit takes
48,001\,$\mu$m$^2$, i.e., 0.60\% of a core.

\vspace{1pt}
\noindent \textbf{Benefits.} In SVT-AV1, the unit speeds up the
scoring by 3.51$\times$. x86 offers instructions only
for SAD~\cite{intel26architectures}, prior SATD hardware hard-wires its
algorithm and block sizes~\cite{partanen21high}, and video chips run encoders in fixed pipelines~\cite{ranganathan21warehouse}. The
unit serves all three encoders, across H.264 and AV1.

\subsection{AI Agents}
\label{sec:case-agents}

For AI agents, \system{} discovers the \emph{Escape Length} and
\emph{Escape Copy} families, which run back to back in the JSON
string encoder of CPython, and which we implement as an ISA
extension and a near-core unit, respectively. AI agents exchange
messages with models and tools as JSON. Escape Length computes the
length of a string after escaping, adding 1, 2, or 6 bytes per character
under an overflow guard. Escape Copy then copies the characters
that need no escaping, and stops at the first one that does.
Escape Length has three members in the ASCII and Unicode encoders,
and Escape Copy has five, also in the UTF-8 decoder.
Their code spans four function names and 29 dataflow graphs, and takes 11.5\% of
the cycles of MiniSWE.

\begin{figure}[h]
\centering
\includegraphics[width=0.85\columnwidth]{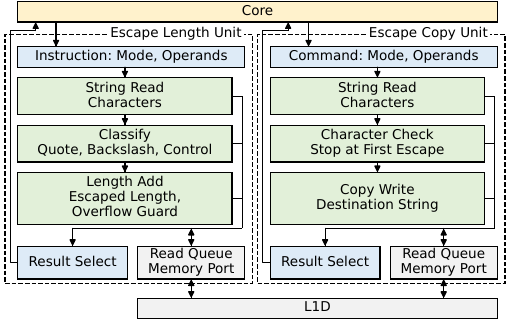}
\caption{Escape Length unit (left) and Escape Copy unit (right) for AI agents.}
\label{fig:case-agents}
\end{figure}

\begin{figure*}[t]
\centering
\includegraphics[width=\textwidth]{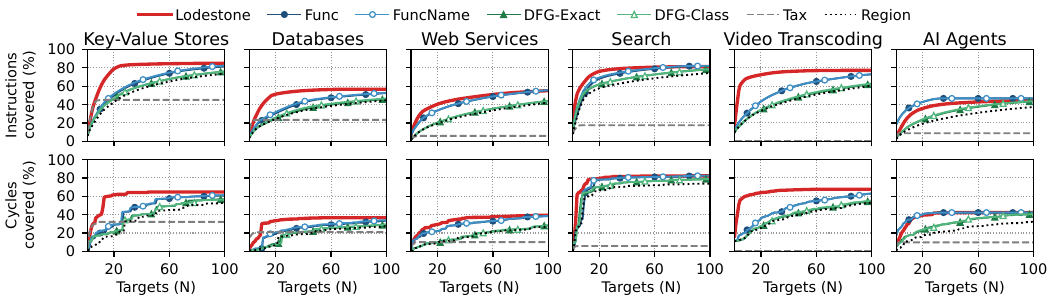}
\caption{Share of the instructions (top) and cycles (bottom) that the top N targets of each method cover in each category.}
\label{fig:coverage}
\end{figure*}

\vspace{1pt}
\noindent \textbf{How Two Units Serve Them.}
Figure~\ref{fig:case-agents} shows the units. Both read several
characters per cycle from the L1 data cache, and the mode selects
the width of the characters. In Escape Length, a classifier marks
quotes, backslashes, and control characters, and an adder
accumulates the escaped length. In Escape Copy, a character check
finds the first character that needs escaping, and the unit copies
the characters before it. They take 1,104 and
3,080\,$\mu$m$^2$, respectively.

\vspace{1pt}
\noindent \textbf{Benefits.} MiniSWE encodes its entire conversation
as JSON on every call to the model. In MiniSWE, the two units speed
up their code by 4.32$\times$ and 6.11$\times$, and together contribute
10.3 percentage points of the 12.8\% end-to-end speedup. In
contrast, prior serialization hardware targets Protocol Buffers and
Thrift~\cite{karandikar21hardware, pourhabibi20optimus}, and no
instruction set classifies characters by JSON rules.

\section{Quantitative Evaluation}
\label{sec:evaluation}

Section~\ref{sec:usecases} presented individual families in depth.
In this section, we evaluate \system{} quantitatively across all workloads: its coverage against the other methods,
the speedup and area of the 30 families that we implement in hardware, the effect of the review,
the stability of the sampling, and the cost of discovery.

\vspace{1pt}
\noindent \textbf{Discovery Coverage.}
\label{sec:eval-coverage}
Figure~\ref{fig:coverage} shows the share of the instructions and cycles that each method covers with N targets.
With N=20, \system{} covers the most instructions and cycles in all categories except AI agents. On average,
it covers 59.1\% of the instructions, i.e., 13.7, 23.6, and 42.3 percentage points more than \emph{FuncName},
\emph{DFG-Class}, and \emph{Tax}, respectively, and it leads by similar margins in cycles.
Even with N=100, \system{} leads \emph{FuncName} by 1.2 points and \emph{DFG-Class} by 8.0 points in instructions.
Hence, with the few targets that designers can afford to implement in hardware, \system{} covers much more of the
execution than 
other techniques. 

\begin{figure*}[t]
\centering
\includegraphics[width=\textwidth]{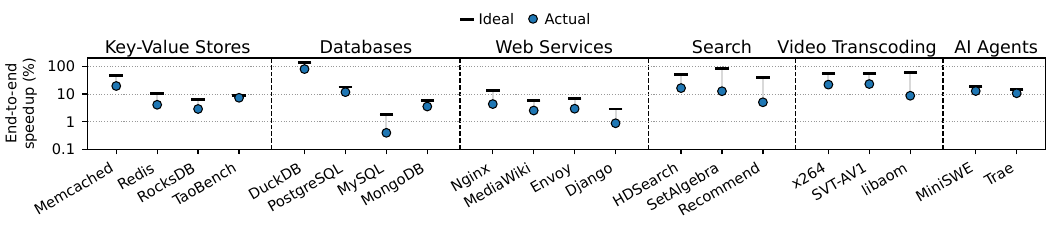}
\caption{End-to-end speedup of each workload with the five units of its category, and the ideal speedup.}
\label{fig:e2e}
\end{figure*}

\vspace{1pt}
\noindent \textbf{Limitations of the Other Methods.} The other methods cannot merge the members of a family.
\emph{Func} and \emph{FuncName} cannot merge code in functions with different names, and \emph{DFG-Exact} and
\emph{DFG-Class} cannot merge code with different dataflow graphs. As we saw, most of the code of a typical family lies
outside its largest function name and dataflow graph. The gap is largest in
video transcoding, where a family has a median of 40 members. There, with N=20, \system{} covers at
least 28.1 points more instructions and cycles than \emph{FuncName} and \emph{DFG-Class}. Conversely,
a function can mix several computations, whereas a family is one computation with a precise definition.
In AI agents, \emph{Func} and \emph{FuncName} cover more instructions than \system{} with N=20,
mainly because their top target is the interpreter loop of CPython. This loop mixes many computations,
so one hardware unit cannot implement it. Finally, the categories of \emph{Tax} are fixed in advance,
so its coverage stops growing after a few targets. Importantly, without a predefined list, \system{} finds
families in all 11 tax categories. For example, data movement has families in all six workload categories,
and memory allocation and hashing in five.

\vspace{1pt}
\noindent \textbf{End-to-End Speedup.}
\label{sec:eval-perf}
Figure~\ref{fig:e2e} shows the end-to-end speedup of each workload with the five units of its category.
On average, the speedup is 12.6\% across the 20 workloads, and 80.2\% in DuckDB.
Databases gain the most, with 24.0\% on average, followed by video transcoding with 17.9\%.
Among the units, Hash Probe speeds up its code the most, by 6.23$\times$ in DuckDB, followed by Escape Copy with
6.11$\times$ in MiniSWE and Hash Mixing with 5.50$\times$ in TaoBench.
The Ideal markers show the speedup if the five families took no time at all.
However, the ideal case also removes the memory accesses of the code, which any unit must still perform.
Hence, the actual speedups reach a median of 41.3\% of the ideal ones.

\begin{figure}[t]
\centering
\includegraphics[width=\columnwidth]{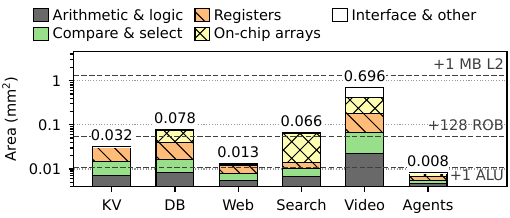}
\caption{Area of the five units of each category.
Dashed lines are the area that common core enhancements add.}
\label{fig:area}
\end{figure}

\vspace{1pt}
\noindent \textbf{Area.} Figure~\ref{fig:area} shows the area of the five units of each category, split by kind of logic, together with the area that three common core enhancements add. In all categories except video transcoding, the five units take at most 0.078\,mm$^2$, i.e., 0.97\% of an Emerald Rapids core. Even the five video units, the largest set, take 0.70\,mm$^2$, less than the 1.32\,mm$^2$ that 1\,MB more L2 adds. Yet, the units give a 12.6\% speedup on average, while one more ALU, 128 more ROB entries, or 1\,MB more L2 gives at most 1.4\%.

\begin{figure}[t]
\centering
\includegraphics[width=\columnwidth]{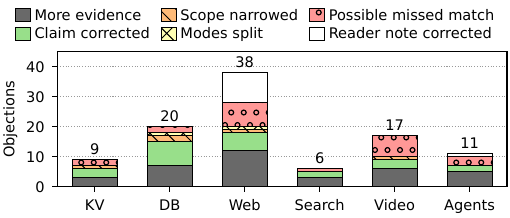}
\vspace{-7mm}
\caption{Objections raised in the agentic AI pipeline.}
\label{fig:review}
\vspace{-2mm}
\end{figure}

\vspace{1pt}
\noindent \textbf{Effect of the Review.}
\label{sec:eval-roles}
Two reviewers check the conclusions of the Matcher, and each catches a different kind of mistake.
Figure~\ref{fig:review} shows the objections that the agents raise in the pipeline of each category.
The Match Reviewer catches false claims, and its objections change the claims of 58 families.
For example, for Cursor Advance,
the Matcher claims that the code clears two fields, but the code only tests them. The Gap Reviewer catches missed
members, and adds 637 members to 18 families. For example, it adds 365 of the 369 members of Symbol Rate Cost,
which extends the family from one video encoder to all three.
The Matcher also corrects 11 Reader notes. Overall, the review changes 67 of the 254 families,
including 13 of the 30 families that we implement in hardware. After the review, all 2,506 members match the
reference implementation of their family on every checked execution, and no member fails.

\begin{figure}[t]
\centering
\includegraphics[width=\columnwidth]{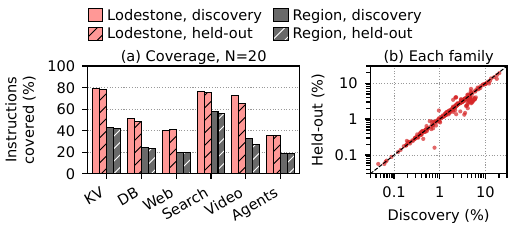}
\vspace{-7mm}
\caption{Instruction coverage on the discovery and held-out windows: (a) of the top 20 targets of \system{} and \emph{Region} in each category, and (b) of each family of \system{}.}
\label{fig:stability}
\end{figure}

\vspace{1pt}
\noindent \textbf{Sampling Stability.}
\label{sec:eval-stability}
\system{} discovers families on a sample of windows, so the families must also cover the executions outside this sample. Figure~\ref{fig:stability}(a) shows the instruction coverage of the N=20 targets of \system{} and \emph{Region} on the discovery and held-out windows. On average, the targets of \system{} cover 59.1\% of the instructions in the discovery windows and 57.3\% in the held-out windows. The largest drop is in video transcoding, from 72.4\% to 65.2\%, where \emph{Region} also drops by 5.6 points. Figure~\ref{fig:stability}(b) shows the coverage of each of the 254 families on the two sets of windows. The coverage of a family differs by a median of 0.10 points, and by at most 0.5 points for 194 families. Hence, the families of \system{} also cover new executions.

\begin{figure}[t]
\centering
\includegraphics[width=\columnwidth]{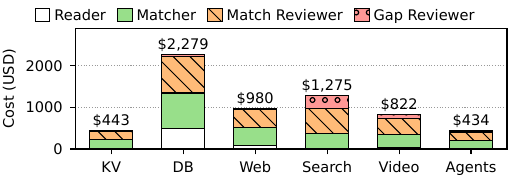}
\vspace{-7mm}
\caption{Cost of one discovery run of each category.}
\label{fig:cost}
\vspace{-4mm}
\end{figure}

\vspace{1pt}
\noindent \textbf{Discovery Cost.}
\label{sec:eval-cost}
Figure~\ref{fig:cost} shows the cost of one discovery run for each category, split by agent role,
at the API price of GPT-6 Astra~\cite{openai26gpt6}. A run costs \$434 to \$2,279 per category,
and the six runs cost \$6,233 in total (\$24.5 per family). The Matcher and Match Reviewer account
for 37.3\% and 43.0\% of the total cost, respectively. 
Discovery costs little compared to the engineering effort of hardware design, and we can rerun it whenever the workloads change.

\section{\system{} with Production Workloads}
\label{sec:production}

To test whether \system{} scales to industrial code, we run it on five large-scale production services from the
fleet of a hyperscaler. These services are proprietary C++ binaries on x86 servers, in two categories:
Category~1 with two services and Category~2 with three.
We run 
with the setup described in Section~\ref{sec:methodology}, and we use an internal LLM.
Since the traces are proprietary, we do not simulate hardware for them, and report coverage and families
instead of speedups.

\vspace{1pt}
\noindent \textbf{Coverage.} Table~\ref{tab:production} shows the instruction coverage of each method.
With N=10 and N=20, \system{} covers the most instructions in both categories.
With N=20, it covers 43.6\% and 24.7\%, 8.1 and 11.0 percentage points more than \emph{Region},
and 6.7 and 9.1 points more than \emph{DFG-Class}. Its lead over \emph{FuncName} is smaller, 0.5 and 2.4 points.
However, a \emph{FuncName} target is a whole C++ function, which mixes control flow and glue logic that one
hardware unit cannot implement, whereas a \system{} family is one computation. In Category~2, whose larger code
footprint spreads the execution more thinly, \system{} needs 12 targets to cover 20\% of the instructions. 
\emph{FuncName}, \emph{Func}, and \emph{Region} need 15, 21, and 53 targets, respectively.

\begin{table}[h]
\vspace{-2mm}
\centering
\footnotesize
\caption{Instruction coverage (\%) of the top N targets of each method in two categories of production services.}
\vspace{-4mm}
\label{tab:production}
\setlength{\tabcolsep}{4pt}
\begin{tabular}{@{}lrrrrrr@{}}
\toprule
 & \multicolumn{3}{c}{Category 1 (2 services)} & \multicolumn{3}{c}{Category 2 (3 services)} \\
\cmidrule(lr){2-4} \cmidrule(lr){5-7}
Method & N=10 & N=20 & N=100 & N=10 & N=20 & N=100 \\
\midrule
\system{} & 35.4 & 43.6 & 52.7 & 19.0 & 24.7 & 31.9 \\
\emph{FuncName} & 34.4 & 43.1 & 52.8 & 17.1 & 22.3 & 31.9 \\
\emph{Func} & 33.4 & 41.2 & 52.5 & 14.1 & 19.6 & 31.4 \\
\emph{DFG-Class} & 29.5 & 36.9 & 49.8 & 11.6 & 15.6 & 26.7 \\
\emph{DFG-Exact} & 29.4 & 36.1 & 49.6 & 10.6 & 14.6 & 25.8 \\
\emph{Region} & 29.2 & 35.5 & 48.6 & 9.8 & 13.7 & 24.9 \\
\emph{Tax} & 26.6 & 26.6 & 26.6 & 14.6 & 14.6 & 14.6 \\
\bottomrule
\end{tabular}
\vspace{-2mm}
\end{table}

\vspace{1pt}
\noindent \textbf{Discovered Opportunities.} The top 10  \system{} 
families cover 35.4\% and 19.0\% of the instructions in
the two categories. In Category~1, the top family is block encryption, which unifies distinct routines of both
services and covers 7.5\% of the instructions alone. The next family, entropy decompression, covers 7.3\% in an
assembly routine that has no symbol at all, so grouping by symbol name or tax category cannot even name it.
Cryptographic and non-cryptographic hashing add 5.4\% and 2.8\%. Category~1 also has workload-specific families,
such as cache eviction and three families that score cache entries, which cover 8.6\% together. In Category~2,
memory allocation and deallocation span all three services and cover 4.3\% and 2.7\% of the instructions.
The allocation family unifies allocator copies that the compiler cloned under different symbols.
Hash-table lookup and string copy (inlined into different callers), and serialization framing (across four specialized handlers) span all three services.

Of the top 10 families in the two categories, 4 and 5 span several services.
Further, 3 and 7 of them unify code that the other methods split across separate or missing symbols,
due to inlining into different callers, compiler-cloned entry points, template specialization, and
stripped assembly routines. These families also match the computations that \system{} finds in the
open-source workloads (hashing, hash-table probing, memory allocation, and data movement).
Thus, even in production binaries, the code of one computation is fragmented
across services and functions. By gathering code into families, \system{} exposes
specialization opportunities in production services that other techniques split or miss.

\section{Related Work}
\label{sec:related}

\noindent \textbf{Analyzing Datacenter Code.} Fleet-wide profiling attributes cycles to leaf functions and
aggregates them into datacenter tax categories: memory allocation, serialization, hashing, and compression~\cite{kanev15profiling, gonzalez23profiling}, and models which of these overheads merit offloading~\cite{sriraman20accelerometer}.
Other studies characterize datacenter workloads, including AI agents~\cite{raj25towards}, and build benchmark suites from them~\cite{ferdman12clearing, sriraman18usuite, gan19open, su25dcperf}.
Binary analysis records control flow at the level of basic blocks~\cite{bruening03dynamorio}, including in profiles sampled across production fleets~\cite{chen16autofdo},
and uses it to reorder code~\cite{panchenko19bolt, shen23propeller} or locate front-end stalls~\cite{ayers19asmdb}.
These works analyze where code runs and stalls, and attribute it to functions or predefined categories.
\system{} analyzes what hot regions compute, and gathers the code of one computation across functions.

\vspace{1pt}
\noindent \textbf{Datacenter Accelerators.} Prior work accelerates datacenter tax, such as serialization~\cite{karandikar21hardware}, remote procedure calls~\cite{lazarev21dagger}, compression~\cite{abali20data}, memory allocation~\cite{kanev17mallacc}, and data movement~\cite{kamath24mc2}, and
orchestrates ensembles of such accelerators~\cite{stojkovic26accelflow}.
Other accelerators target application logic, such as key-value stores~\cite{li17kvdirect}, databases~\cite{kocberber13meet, zhang19leveraging, yuan21qei}, dynamic language runtimes~\cite{gope17architectural}, search engines~\cite{heo20iiu}, tree traversal~\cite{ha24generalizing, barnes24extending, lee24terminus}, and video transcoding~\cite{ranganathan21warehouse, partanen21high}.
Processors and hyperscalers also deploy on-chip accelerators~\cite{intel22technical, bilgin25announcing}, custom chips for AI~\cite{google26introduction, krishnamurthy26mtia, guthrie26maia}, and ISA extensions~\cite{intel26architectures}.
All these designs target computations that experts chose in advance, whereas \system{} automatically
finds fine-grained specialization opportunities.

\vspace{1pt}
\noindent \textbf{Automated Target Discovery.} Instruction-set extension methods enumerate hot code dataflow subgraphs~\cite{atasu03automatic} and group isomorphic ones into custom instructions~\cite{rezunov25cidre, hagerclukas25arise}. Others synthesize hot code into specialized cores, datapaths, or inline accelerators~\cite{venkatesh10c-cores, trilla21novia, lysecky06warp}. QsCores merges structurally similar code into one core~\cite{venkatesh11qscores}, Potluck shares modules with the same names across accelerator designs~\cite{chen26potluck}, and ISAMORE rewrites programs with equivalence rules to find instructions that recur across them~\cite{xiao26isamore}. These methods merge code when its structure or name matches or when rewrite rules prove it equivalent. The code of a computation that differs in algorithms, control flows, or data layouts remains fragmented. Algorithm recognition can match such code, but only for computations named in advance~\cite{woodruff22bind,martinez24code}.
\system{} groups code by what it computes, without a predefined list of target computations.

\vspace{1pt}
\noindent \textbf{LLMs for Systems.} LLMs increasingly design and optimize systems, such as job schedulers and accelerator circuits~\cite{novikov25alphaevolve}, cache replacement~\cite{gupta26archagent}, prefetching~\cite{gonzalez26archagentv2}, compute kernels~\cite{liao26kernelevolve}, control-plane policies~\cite{lin26atumai}, benchmarks~\cite{andrade26fidelio}, and EDA scripts~\cite{liu23chipnemo}. Closest to our work, A3D uses agents to generate accelerators for the bottleneck functions of an application~\cite{nallathambi26a3d}, and HSCO-Bench evaluates agents on this task~\cite{tsai26hscobench}. In contrast, \system{} discovers computations shared across independently developed binaries, so that one hardware design serves all members of a family.

\section{Conclusion}
\label{sec:conclusion}

Hardware specialization can improve datacenter efficiency, but the computations that justify it are
often fragmented across many different pieces of code.
We presented \system{}, an agentic framework that discovers such computations by reasoning about what the
executed code computes.
Thirty hardware specialization opportunities discovered by \system{} speed up the workloads by 12.6\% on average and by up to 80.2\%.

%%%%%%% -- PAPER CONTENT ENDS -- %%%%%%%%

%%%%%%%%% -- BIB STYLE AND FILE -- %%%%%%%%
\bibliographystyle{ACM-Reference-Format}
\bibliography{refs}
%%%%%%%%%%%%%%%%%%%%%%%%%%%%%%%%%%%%

\end{document}